\documentclass[journal]{IEEEtran}
\usepackage{cite}
\usepackage{amsmath,amssymb,amsfonts}
\usepackage{algorithmic}
\usepackage{graphicx}
\usepackage{textcomp}
\begin{document}
\title{Latent Diffusion Model for Medical Image Standardization and Enhancement}
\author{Md Selim, Jie Zhang, Faraneh Fathi, Michael A. Brooks, Ge Wang, Guoqiang Yu, and Jin Chen 
\thanks{This research is supported by NIH (grant no. R21 CA231911, R01 EB028792, R01 HD101508, R41 NS122722, R42 MH135825) and Kentucky Lung Cancer Research (grant no. KLCR-3048113817). }
\thanks{Md Selim is an ORISE Fellow with the US Food and Drug Administration. This work is conducted during his graduate student status at the Department of Computer Science and the Institute for Biomedical Informatics, Lexington, KY 40503 USA (e-mail: md.selim@uky.edu). }
\thanks{Jie Zhang is with the Department of Radiology, University of Kentucky, Lexington, KY, USA (e-mail: jnzh222@uky.edu).}
\thanks{Faraneh Fathi is with the Department of Biomedical Engineering, University of Kentucky, Lexington, KY, USA (e-mail: faraneh.fathi@uky.edu).}
\thanks{Michael A. Brooks is with the the Department of Radiology, University of Kentucky, Lexington, KY, USA (e-mail: mabroo3@uky.edu).}
\thanks{Ge Wang is with the Biomedical Imaging Center, Rensselaer Polytechnic Institute, Troy, NY, USA (e-mail: wangg6@rpi.edu).}
\thanks{Guoqiang Yu is with Department of Biomedical Engineering, University of Kentucky, Lexington, KY, USA (e-mail: gyu2@uky.edu).}
\thanks{Jin Chen is with the Department of Medicine and Informatics Institute, University of Alabama at Birmingham, AL, USA (e-mail: jchen5@uab.edu).}
}

\maketitle

\begin{abstract}
Computed tomography (CT) serves as an effective tool for lung cancer screening, diagnosis, treatment, and prognosis, providing a rich source of features to quantify temporal and spatial tumor changes.
Nonetheless, the diversity of CT scanners and customized acquisition protocols can introduce significant inconsistencies in texture features, even when assessing the same patient. This variability poses a fundamental challenge for subsequent research that relies on consistent image features.
Existing CT image standardization models predominantly utilize GAN-based supervised or semi-supervised learning, but their performance remains limited. We present DiffusionCT, an innovative score-based DDPM model that operates in the latent space to transform disparate non-standard distributions into a standardized form. 
The architecture comprises a U-Net-based encoder-decoder, augmented by a DDPM model integrated at the bottleneck position. First, the encoder-decoder is trained independently, without embedding DDPM, to capture the latent representation of the input data. Second, the latent DDPM model is trained while keeping the encoder-decoder parameters fixed. Finally, the decoder uses the transformed latent representation to generate a standardized CT image, providing a more consistent basis for downstream analysis.
Empirical tests on patient CT images indicate notable improvements in image standardization using DiffusionCT. Additionally, the model significantly reduces image noise in SPAD images, further validating the effectiveness of DiffusionCT for advanced imaging tasks.
\end{abstract}

\begin{IEEEkeywords}
CT imaging, image standardization, image synthesis, diffusion
\end{IEEEkeywords}

\section{Introduction}\label{sec:intro}
\IEEEPARstart{L}{ung} cancer is the leading cause of cancer death and is among the most prevalent types of cancer for both men and women in the United States~\cite{collins2017letter}. The overall 5-year survival rate for non-small cell lung cancer (NSCLC) is approximately 19\%. Computed tomography (CT) imaging plays a critical role in the early diagnosis of lung cancer and aids in defining tumor characteristics for better treatment outcomes~\cite{de2014benefits,ravanelli2013texture}.  Texture features extracted from CT images may quantify spatial and temporal variations in tumor architecture and function, allowing for the determination of intra-tumor evolution~\cite{ardila2019end,song2017using}. However, the use of CT scanners from different vendors, each with its own customized acquisition protocols, introduces significant variability in the texture features of images, even when observing the same patient. This inconsistency presents a substantial challenge for conducting large-scale studies across multiple sites~\cite{lu2020deep}. The absence of standardized radiomics consequently hampers the reliability and effectiveness of downstream clinical tasks.

\begin{figure}[!bt]
\centering
\includegraphics[width=\columnwidth]{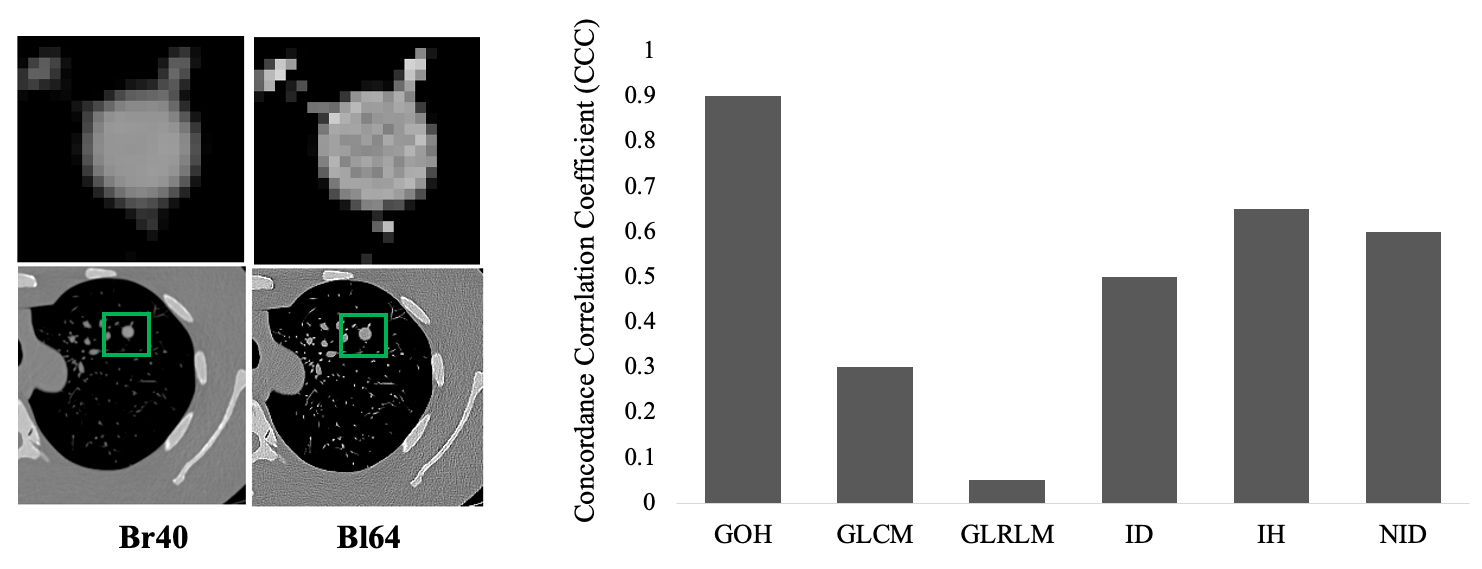}
\caption{{\bf Discrepancy of tumor image features caused by different imaging protocols.} The same lungman chest phantom was scanned using the same scanner. CT images were acquired using two different image reconstruction kernels accordingly, as indicated by the texts at the bottom of the images. In the images on the left side, a tumor is marked with green rectangles (the top row is the zoomed-in tumor regions respectively). The histogram on the right side showed the feature variance between these two tumors in terms of CCC. The observed differences in the tumor images may have significant implications on the promise of large-scale radiomic studies.  } \label{fig:senario} 
\end{figure}

Inconsistency in radiomic features, including texture, shape, and intensity, is a known issue when images are captured using different scanners from various vendors or even with different acquisition protocols on the same scanner~\cite{berenguer2018radiomics,hunter2013high}. This inconsistency, both within a single scanner using various settings and across different scanners using similar settings, presents a persistent challenge that needs to be addressed. 
%
%
Figure~\ref{fig:senario} shows an example of the impact of non-standard CT imaging acquisition protocols on radiomic features. 
A lungman chest phantom, equipped with three artificial tumors, was scanned using Siemens CT scanners. The resulting images were reconstructed using two different Siemens reconstruction kernels Bl64 and Br40. The visual characteristics and radiomic features of the tumors varied notably in images generated with different reconstruction kernels.  
%

Developing a universal CT image acquisition standard has been suggested as a potential solution. 
However, implementing this standard would require substantial modifications to existing CT imaging protocols, and could potentially narrow the scope of applications for the modality~\cite{paul2012relationships,gierada2010effects}. Given these constraints, alternative approaches are needed to address the issue of radiomic feature discrepancies in CT images.
%

Recent advancement has been made to address the CT radiomic feature variability problem. One promising solution is to develop a post-processing framework capable of standardizing and normalizing existing CT images while preserving anatomic details~\cite{ours_aamp,cohen2012radiosity,selim2020stan,selim2021cross,selim2021radiomic}. Our research indicates that this approach allows for the extraction of reliable and consistent features from standardized images, facilitating accurate downstream analysis, and ultimately leading to improved diagnosis, treatment, and prognosis of lung cancer. Deep learning algorithms for image standardization are particularly promising for harmonizing CT images taken with diverse parameters on the same scanner~\cite{selim2020stan}. 
%
%
It is, nevertheless, important to recognize that current solutions exhibit limitations, particularly in image texture synthesis and maintaining structural integrity. All these can adversely affect the performance of subsequent analyses, thereby impeding the development of dependable and consistent features that are crucial for enhancing lung cancer diagnosis, treatment, and prognosis.
Continued research is crucial for advancing algorithms to address these challenges and augment the performance of CT image standardization. Progress in this domain has the potential to substantially improve the quality of medical imaging, contributing to the development of more effective strategies for combating lung cancer.

\begin{figure*}[bt!]
\centering
\includegraphics[width=\textwidth]{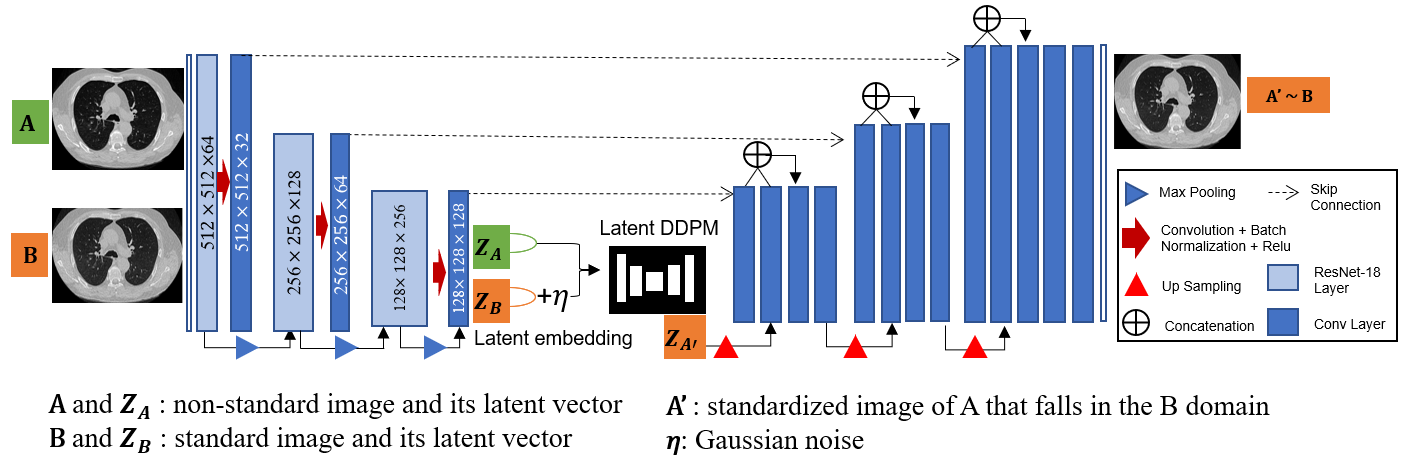}
\caption{Overview of a score-based DDPM pipeline for intra-scanner standardization. Given an image pair (A, B) where A and B are non-standard and the corresponding standard images, the model aims to synthesize a new image $A'$ in domain $B$. The representation learning component learns encoded latent representations of CT images using a ResNet-18-based encoder-decoder structure. The target-specific latent-space mapping component is designed for standard image synthesis. It contains a DDPM model for latent space mapping. $Z_A$ is the latent vector of non-standard image A; $Z_B$ is the latent vector of standard image B; $Z_{A'}$ is the standardized latent vector of image A, and $\eta$ is Gaussian noise. } \label{fig:diff_model}
\end{figure*}

Compared to the state-of-the-art generative adversarial networks (GAN) and variational auto-encoders (VAE) algorithms, score-based denoising diffusion probabilistic models (DDPM)~\cite{ho2020denoising} shows superior performance in image standardization. DDPM learns a Markov chain to gradually convert a simple distribution, such as isotropic Gaussian, into a target data distribution. It consists of two processes: (1) a fixed forward diffusion process that gradually adds noise to an image when sequentially sampling latent variables of the same dimensionality and (2) a learned reverse denoising diffusion process, where a neural network (such as U-Net) is trained to gradually denoise an image starting from a pure noise realization. DDPM and its variants have attracted a surge of attention since 2020, resulting in key advances in continuous data modeling, such as image generation~\cite{ho2020denoising}, super-resolution~\cite{saharia2022image}, and image-to-image translation~\cite{saharia2022palette}. More recently, conditional DDPM has shown remarkable performance in conditional image generation~\cite{yang2018low}. In parallel, latent DDPM enables generating image embedding in a low-dimensional latent space. 


Building on recent advancements in DDPM, this study introduces DiffusionCT, an innovative solution for CT image standardization. The architecture of DiffusionCT combines an encoder-decoder network with a latent conditional DDPM, as illustrated in Fig~\ref{fig:diff_model}. The encoder-decoder network maps the input CT image to a low-dimensional latent representation. The DDPM then models the conditional probability distribution of the latent representation to synthesize a standard image. This innovative framework aims to address the current limitations in CT image standardization and contribute to more reliable medical imaging for lung cancer management. Notably, DiffusionCT preserves the original structure of the CT image while effectively standardizing texture.


Additionally, we demonstrate that the capabilities of DiffusionCT can be extended beyond standardization to include effective noise reduction in medical images. As part of our case study, we applied DiffusionCT to the 2D mapping of cerebral blood flow (CBF) images at different depths of the head captured with the time-resolved laser speckle contrast imaging (TR-LSCI) technology. DiffusionCT successfully denoised the blurry depth images, thereby recovering high-quality CBF maps. This extended capability broadens the tool's applicability across diverse medical imaging tasks and further solidifies its potential in enhancing diagnostic and treatment strategies across a range of medical conditions.

%

\section{Background} \label{sec_bcg}

\subsection{CT Image Acquisition and Reconstruction Parameters}
CT images are typically acquired by setting several parameters, such as kilovoltage peak (kVp), Pitch, milliamperes-second (mAs), reconstruction field Of view (FOV), slice thickness, reconstruction kernels, etc. 
Varying the settings of CT image acquisition and reconstruction parameters and the selection of different CT scanners may subsequently alter radiomic features extracted from the images. For instance, in Figure~\ref{fig:senario}, the Br40 kernel produces a smoother image, while the Bl64 kernel results in a sharper image. These differing texture patterns will yield distinct radiomic features, complicating subsequent clinical tasks.

\subsection{Radiomic Features} 
Radiology employs sophisticated non-invasive imaging technologies for the diagnosis and treatment of various diseases. Crucial to tumor characterization are the image features extracted from radiological images using mathematical and statistical models~\cite{yip2016applications}. 
Among these features, radiomic features provide insight into the cellular and genetic levels of phenotypic patterns hidden from the naked eyes~\cite{ yang2011quantifying, basu2011evolving, yip2016applications}. 
Radiomic features can be categorized into six classes: Gradient Oriented Histogram (GOH), Gray Level Co-occurrence Matrix (GLCM), Gray Level Run Length Matrix (GLRLM), Intensity Direct (ID), Intensity Histogram (IH), and Neighbor Intensity Difference (NID). 

Utilizing radiomic features offers considerable potential to capture tumor heterogeneity and detailed phenotypic information. However, the efficacy of radiomic studies, especially in the context of extensive cross-institutional collaborations, is significantly hindered by the lack of standardization in medical image acquisition practices~\cite{hunter2013high, berenguer2018radiomics}. 


\subsection{CT Image Standardization Approaches} 
In general, There are two types of CT image standardization approaches, each serving distinct purposes and contingent upon data availability. 
The first category, known as intra-scanner image standardization, necessitates the availability of paired image data~\cite{selim2020stan}. In this scenario, two images constructed from the same scan but employing different reconstruction kernels constitute an image pair, where the source image refers to the image constructed with the non-standard kernel (e.g., Siemens Br40), and the target image is constructed  using the standard kernel (e.g., Siemens Bl64). Given paired image data as the training data, a machine learning model is trained to convert source images to target images. 
The second category for CT image standardization encompasses models devised for cross-scanner image standardization, which eliminates the need for paired image data~\cite{selim2021cross}. In this setting, images are not required to be matched; rather, images acquired with different protocols are stored separately. 

Acquiring paired training data is straightforward, though it is predominantly confined to a single scanner. In large-scale radiomic studies, the need for standardization is more pronounced in cross-vendor scenarios, which cannot be accomplished by utilizing models from the first category. To address the issue of cross-vendor image standardization, models in the second category mitigate the requirement for paired images, albeit at the cost of reduced performance.

%
Liang et al~\cite{liang2019ganai} developed a CT image standardization model, denoted as GANai, based on conditional Generative Adversarial Network (cGAN)~\cite{pix2pix}. A new alternative training strategy was designed to effectively learn the data distribution. GANai achieved better performance in comparison to cGAN and the traditional histogram matching approach~\cite{gonzalez2012digital}. However, GANai primarily focuses on the less challenging task of image patch synthesis  rather than addressing the entire DICOM image synthesis problem.

Selim et al~\cite{selim2020stan} introduced another cGAN-based CT image standardization model, denoted as STAN-CT. In STAN-CT, a complete pipeline for systematic CT image standardization was constructed. Also, a new loss function was devised to account for two constraints, i.e., latent space loss and feature space loss. The latent space loss is adopted for the generator to establish a one-to-one mapping between  standard and synthesized images. The feature space loss is utilized by the discriminator to critique the texture features of the standard and the synthesized images. Nevertheless, STAN-CT was limited by the limited availability of training data and was evaluated at the image patch level on a limited number of texture features, utilizing only a single evaluation criterion. 

RadiomicGAN, another GAN-based model, incorporates a transfer learning approach to address the data limitation issue~\cite{selim2021radiomic}. The model is designed using a pre-trained VGG network. A novel training technique called window training is implemented to reconcile the pixel intensity disparity between the natural image domain and the CT imaging domain.  Experimental results indicated that RadiomicGAN outperformed both STAN-CT and GANai.

For cross-scanner image standardization, a model termed CVH-CT was developed~\cite{selim2021cross}. CVH-CT aims to standardize images between scanners from different manufacturers, such as Siemens and GE. The generator of CVH-CT employs a self-attention mechanism for learning scanner-related information. A VGG feature-based domain loss is utilized to extract texture properties from unpaired image data, enabling the learning of scanner-based texture distributions. Experimental results show that, in comparison to CycleGAN~\cite{CycleGAN2017}, CVH-CT enhanced feature discrepancy in the synthesized images, but its performance is not significantly improved when compared with models trained within the intra-scanner domain. 

UDA-CT, a recently developed deep learning model for CT image standardization, demonstrates a departure from previous methods by incorporating both paired and unpaired images, rendering it more flexible and robust~\cite{selim2022udact}. UDA-CT effectively learns a mapping from all non-standard distributions to the standard distribution, thereby enhancing the modeling of the global distribution of all non-standard images. Notably, UDA-CT demonstrates compatible performance in both within-scanner and cross-scanner settings. 


The development of standardization models for CT images has provided a solid foundation for generating stable radiomic features in large-scale studies. However, recent advances in image synthesis using diffusion models have opened up new opportunities for investigating the CT image standardization problem. These models offer a powerful approach for generating high-quality, standardized images from diverse sources, which could greatly improve the accuracy and reliability of radiomic studies. By leveraging the strengths of both standardization and synthesis models, researchers may be able to unlock new insights into the relationship between CT images and disease outcomes.


\section{Method} \label{sec_method}
The structure of DiffusionCT is shown in Fig~\ref{fig:diff_model}, encompassing two major components: the image embedding component and a conditional DDPM in the latent space. The image embedding component employs an encoder-decoder network to translate input CT images to a low-dimensional latent representation. Subsequently, the conditional DDPM models the conditional probability distribution of the latent representation in order to synthesize a standard image. Importantly, DiffusionCT retains the original structure of the input image while effectively standardizing its texture.


DiffusionCT is trained sequentially in three steps. First, in the pre-processing step, the encoder-decoder network is trained with all CT images in the training set, irrespective of whether they are standard or non-standard or whether they are captured using GE or Siemens. This step aims to effectively encode images into a 1-D latent vector, which can reconstruct the original image with minimal information loss. Second, a latent conditional DDPM is trained with image pairs, consisting of a non-standard image and its corresponding standard image. This step enables the DDPM to model the conditional probability distribution of the latent representation, thus facilitating the synthesis of standard images. Finally, all the trained neural networks are combined to standardize new images.

\subsection{Image encoding and decoding}
\label{sec_model}
The image embedding component of DiffusionCT comprises a customized U-Net structured convolutional network, designed to learn a low-dimensional latent representation of input images. The encoder and decoder of the U-Net are asymmetric. 
The encoder uses a pre-trained ResNet-18 with four neural blocks. The first convolutional block consists of the first three ResNet-18 layers. The second block consists of the fourth and fifth layers of ResNet-18. The 3rd, 4th, and 5th blocks of the encoder consist of the corresponding 5th, 6th, and 7th layers of ResNet-18, respectively. The decoder encompasses a five-block convolutional network with up-sampling and several 1D convolutional layers in the last layers. Skip connection is not used within the 1D convolutional layers. 

This novel U-Net is trained with all available images in the training dataset, irrespective of whether they are standard or non-standard, in order to learn a global image encoding. The anatomic loss is adopted to facilitate the learning of structural information within the images. The trained U-Net encodes an input image into a latent low-dimensional representation, and the decoder accepts a latent representation to reconstruct the input image. This step is applicable for both intra-scanner and cross-vendor image standardization. The L2-regularized loss function is adopted for model training. 

\subsection{Conditional latent DDPM}

\begin{figure}[bt!]
\centering
\includegraphics[width=.5\textwidth]{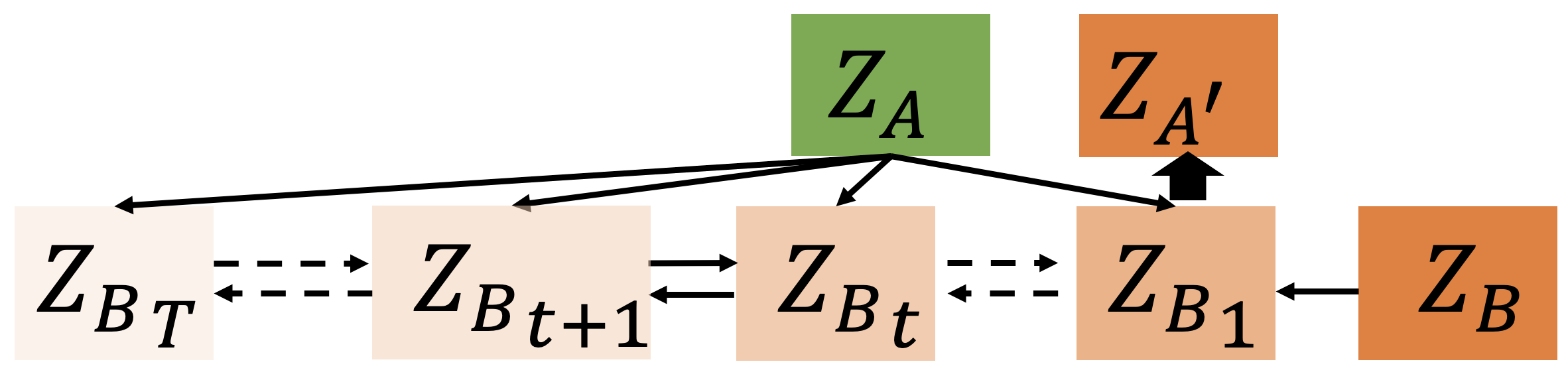}
\caption{\textbf{Conditional latent DDPM} for converting embedding $Z_A$ to $Z_{A'}$ in the B domain. } \label{fig:diff_lat}
\end{figure}

In the context of intra-scanner image standardization, paired image data are provided, consisting of a non-standard image $A$ and the corresponding standard image $B$. Using the previously described trained encoder, latent embeddings $Z_A$ and $Z_B$ are generated from non-standard ($A$) and standard ($B$) images, respectively. As $Z_A$ and $Z_B$ adhere to distinct distributions, a conditional latent DDPM is designed to map the non-standard latent distribution to the standard latent distribution. The encoder-decoder network remains unaltered during diffusion training. A well-trained conditional latent DDPM preserves anatomic details in $Z_A$ while mapping texture details from $Z_A$ to $Z_B$. 

Structure-wise, the conditional latent DDPM includes multiple small steps of diffusion in each training step. In every individual diffusion step, Gaussian noise $\eta$ is added to the latent embedding $Z_B$. All the corrupted $Z_B$  conditioned to $Z_A$ are used to train the conditional latent DDPM described in Figure~\ref{fig:diff_lat}. For a significant large $T$, where $T$ represents the total number of diffusion steps, $\prod_{t-1}^{T}(Z_{B_t}+\eta)$ converges to an isotropic Gaussian distribution. 

The network structure of the conditional latent DDPM is a U-Net, which is trained to predict the added noise $\eta$ from $\prod_{t-1}^{T}(Z_{B_t}+\eta)$. In addition to the standard diffusion loss function (see details at Ho et al~\cite{ho2020denoising}), an L1-loss between the reconstructed and the non-standard embeddings $\mathcal{L} = \mathbb{E}_{t\sim [1,T]} [|\eta_t  - p_\theta(Z_A,Z_{B_t})|]$ is used to update the diffusion model. After training, for each non-standard embedding $Z_A$, the model synthesizes a latent standardized embedding $Z_{A'}$.

\subsection{Model training}
To ensure effective training, we consider a two-step strategy, i.e., representation learning and latent diffusion training. In representation learning, we train the customized U-Net with all training images to learn the latent low-dimensional representation. Specifically, the network introduced in \ref{sec_model} is trained to learn the global data representation of all training images in the latent space. After the encoder and decoder are well trained, they remain fixed, and the latent diffusion model training starts. 
In the latent diffusion training process, we train the proposed conditional latent DDPM introduced in~\ref{sec_model} to map the latent representation of non-standardized images to the standard image domain. 

The trained encoder-decoder network and conditional latent DDPM are integrated for image standardization. A non-standard image $A$ is passed through the trained encoder to convert it into a latent representation $Z_A$. Then, $Z_A$ is passed through the trained conditional latent DDPM to generate $Z_{A'}$, which falls into the standard embedding domain. Finally, $Z_{A'}$ is passed through the trained decoder to synthesize image $A'$ in the standard image domain $B$.

\section{Experimental Results}
DiffusionCT was built using the PyTorch framework. The network weights were randomly initialized. The learning rate was set to $10^{-4}$ with the Adam optimizer. The encode-decode network underwent training for a duration of 20 epochs, followed by an additional 20 epochs dedicated to training the diffusion network.  In total, the model required about 20 hours for complete training from scratch. Once the model was fully trained, it took about 30 seconds to process and synthesize a standardized slice of a DICOM CT image. 

We compared DiffusionCT with five recently developed CT image standardization models, including GANai~\cite{liang2019ganai}, STAN-CT~\cite{selim2020stan}, and, RadiomicGAN~\cite{selim2021radiomic}, CVH-CT~\cite{selim2021cross}, and UDA-CT~\cite{selim2021cross}, as well as the original DDPM and the encoder-decoder network.  
%
To evaluate the model performance, the results were measured using two metrics: the concordance correlation coefficient (CCC) and error rate. These metrics allow for a quantitative evaluation of the effectiveness of the proposed method in achieving CT image standardization while preserving the original texture and structure of the images.

\subsection{Experimental Data} 
%

The training data consist of a total of 9,886 CT image slices from 14 lung cancer patients captured using two different kernels (Br40 and Bl64) and 1mm slice thickness using a Siemens CT Somatom Force scanner at the University of Kentucky Albert B. Chandler Hospital. 
The training data also contain additional 9,900 image slices from a lungman chest phantom scan, with three synthetic tumors inserted. The phantom is scanned using two different kernels (Br40 and Bl64) and two different slice thicknesses of 1.5mm and 3mm using the same scanner. In total, 19,786 CT image slices were used to train DiffusionCT. 
To prepare the testing data, the identical lungman chest phantom was used. The testing data comprised  126 CT image slices acquired using two different kernels (Br40 and Bl64) with a Siemens CT Somatom Force scanner. Notably, despite the commonality of the phantom used in obtaining both training and testing data sets, the acquisition of test data with a 5mm slice thickness results in the disjoint nature of the training and testing data.  
In this experiment, for demonstration purposes, Siemens Bl64 is considered the standard protocol, while Siemens Br40 was regarded as non-standard. Our standardization experiments focus to mitigate reconstruction kernel-related variability.   



\begin{figure*}[!bt]
\centering
\includegraphics[width=\textwidth]{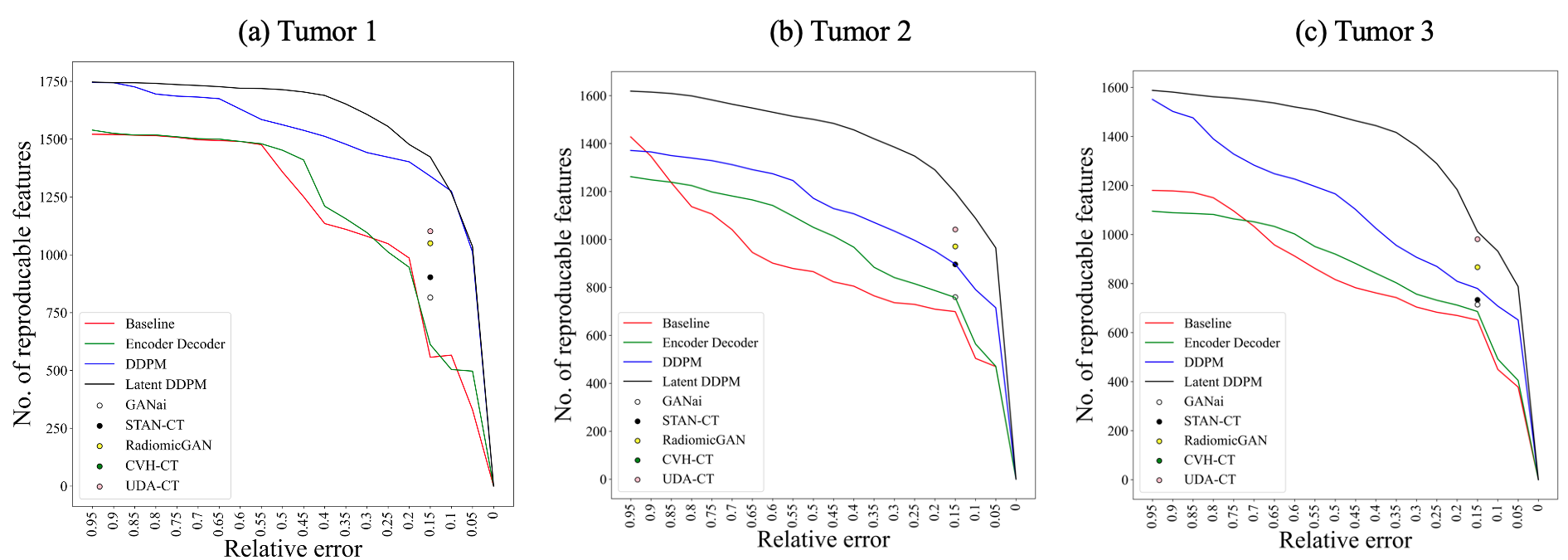}
\caption{\textbf{Total number of reproducible features after Siemens Br40 image synthesis.} Each point on the line represents the total number of reproducible features for the respective error threshold. The existing models' performances are denoted by the circle-shaped points only for the error threshold $RE<0.15$. } \label{fig:se_re}
\end{figure*}

\subsection{Evaluation Metric}
Model performance was evaluated based on lung tumors in the CT images. For each tumor, a total of 1,401 radiomic features, from six feature classes (GOH, GLCM, GLRLM, ID, IH, NID), were extracted using IBEX~\cite{zhang2015ibex}. 
Based on these radiomic features, we evaluated DiffusionCT and all the baseline models using two evaluation metrics, with one-to-one feature comparison and group-wise comparison.  

First, the error rate, defined as the relative difference between a synthesized image and its corresponding standard image regarding a radiomic feature, was utilized to calculate the linear distance between the standard and the synthesized images regarding each individual radiomic feature. the error rate ranges from 0 to 1, and is the lower the better.  
\begin{equation}\label{eq:error_distance}
ErrorRate(s,t) = \frac{|f_t - f_s|}{f_t} \times 100\%
\end{equation}
\noindent where $f_t$ and $f_s$ are the radiomic feature values of the standard and synthesized image, respectively;  and $s$ and $t$ stand for the standard and the synthesized images, respectively.

Usually, a radiomic feature is considered to be reproducible if the synthesized image is more than 85\% similar to the corresponding standard image~\cite{zhao2016reproducibility,choe2019deep}. Mathematically, a radiomic feature is considered reproducible if and only if   $ErrorRate(s,t) < 15\%$. 

Concordance Correlation Coefficient~\cite{lawrence1989concordance} (CCC) was employed to measure the level of similarity between two feature groups~\cite{choe2019deep}. Mathematically, CCC represents the correlation between the standard and the non-standard image features in the radiomic feature class $r$. CCC ranges from -1 to 1, and is the higher the better.  
\begin{equation}\label{eq:ccc}
    CCC(s,t, r) = \frac{2\rho_{s,t,r} \sigma _s \sigma_t}{{\sigma _s}^2 +{\sigma_t}^2 + {( \mu _s - \mu_t )}^2}
\end{equation}
\noindent where $s$ and $t$ stand for the standard and the synthesized images, respectively; $\mu_s$ and $\sigma_s$ (or $\mu_t$ and $\sigma_t$) are the mean and standard deviation of the radiomic features belonging to the same feature class $R$ in a synthesized (or standard) image, respectively; and $\rho_{s,t,r}$ is the Pearson correlation coefficient between $s$ and $t$ regarding a feature class $r$.

\subsection{Results and Discussion}

\begin{table*}[!bt]
 	\centering
 	\caption { The CCC values of the images synthesized using different image standardization models. Each column represents the mean$\pm$std CCC values of lung tumor ROIs for a specific radiomic feature group. }
 	\begin{tabular}{ c  c  c  c c  c  c    }
 		\hline
Feature Class	&	GOH			&	GLCM			&	GLRLM			&	ID			&	IH			&	NID			\\ \hline

Baseline	&	0.90	$\pm$	0.05	&	0.20	$\pm$	0.13	&	0.59	$\pm$	0.13	&	0.33	$\pm$	0.16	&	0.35	$\pm$	0.12	&	0.28	$\pm$	0.15	\\ 
GANai	&	0.95	$\pm$	0.05	&	0.50	$\pm$	0.08	&	0.63	$\pm$	0.12	&	0.59	$\pm$	0.03	&	0.44	$\pm$	0.08	&	0.65	$\pm$	0.10	\\ 
STAN-CT	&	0.95	$\pm$	0.05	&	0.70	$\pm$	0.10	&	0.72	$\pm$	0.15	&	0.75	$\pm$	0.16	&	0.61	$\pm$	0.11	&	0.71	$\pm$	0.05	\\ 
RadiomicGAN	&	1.00	$\pm$	0.00	&	0.80	$\pm$	0.12	&	0.75	$\pm$	0.11	&	0.82	$\pm$	0.08	&	0.72	$\pm$	0.09	&	0.73	$\pm$	0.12	\\ \hline				
Encoder-Decoder	&	1.00	$\pm$	0.00	&	0.38	$\pm$	0.19	&	0.61	$\pm$	0.15	&	0.52	$\pm$	0.11	&	0.39	$\pm$	0.25	&	0.33	$\pm$	0.09	\\ \hline			
DDPM	&	1.00	$\pm$	0.00	&	0.81	$\pm$	0.23	&	\textbf{0.80	$\pm$	0.18}	&	0.85	$\pm$	0.15	&	\textbf{0.77	$\pm$	0.12}	&	0.82	$\pm$	0.13	\\ 
DiffusionCT	&	\textbf{1.00	$\pm$	0.00}	&	\textbf{0.85	$\pm$	0.14}	&	0.79	$\pm$	0.21	&	\textbf{0.89	$\pm$	0.28}	&	0.41	$\pm$	0.05	&	\textbf{0.86	$\pm$	0.18}	\\ \hline	

 	\end{tabular} \label{table:a_dataset}
\end{table*}

In Figure~\ref{fig:se_re}, each point on a line represents the total number of radiomic features on the y-axis whose respective error rate is equal to or smaller than the value specified  on the x-axis.
The red line represents the direct comparison of the input images and the corresponding standard images without using any algorithms. The green, blue, and black lines represent the performance of the encoder-decoder network, DDPM, and DiffusionCT model, respectively. In the literature, the compared models' performances were reported based on a 15\% error rate. In figure~\ref{fig:se_re}, the model performance on $ErrorRate\leq 0.15$ showed that DiffusionCT preserved 64\% and DDPM preserved 58\% more radiomic features than the baseline, comparing to GANai at 20\%, STAN-CT at 32\%, and RadiomicGAN at 51\%. 

Table~\ref{table:a_dataset} shows the CCC scores of six  classes of radiomic features. The performance of the baseline was measured using the input images. In four out of six feature classes, DiffusionCT achieved $CCC >0.85$, clearly outperforming all the compared models. Nevertheless,  DDPM outperformed DiffusionCT and other compared models in two other feature groups. Notably, GLCM and GLRLM together occupy almost 50\% of the total number of radiomic features, and both the DDPM and our DiffusionCT achieved significant performance gains. Also, DDPM had the highest variation on GLCM, indicating conditional DDPM could be more suitable for the image standardization task

\begin{figure*}[!tb]
\centering
\includegraphics[width=.8\textwidth]{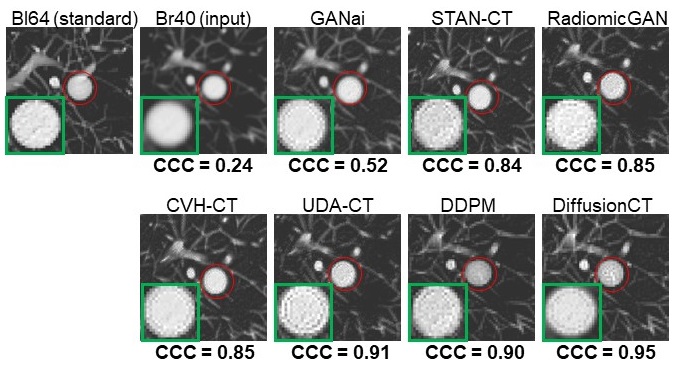}
\caption{CT images synthesized using all compared models in the display window of [-800, 600] HU. The leftmost image is the standard image, and the right bottom is the result of DiffusionCT. Each image contains the same ROI with a tumor marked in a red circle and magnified in the green box. CCC scores of GLCM are displayed at the bottom.} \label{fig:example}
\end{figure*}



Figure~\ref{fig:example} visualizes the results of all compared models on a sample tumor. The input tumor image is observably different from the standard image regarding visual appearances as well as radiomic features. The DiffusionCT-generated image has the highest CCC values regarding GLCM in reference to the standard image and is visually more similar to the standard image than the ones generated by GAN-based models and the vanilla DDPM.

\subsection{Case study on TR-LSCI image denoising}
\begin{figure*}[bt!]
\centering
\includegraphics[width=\textwidth]{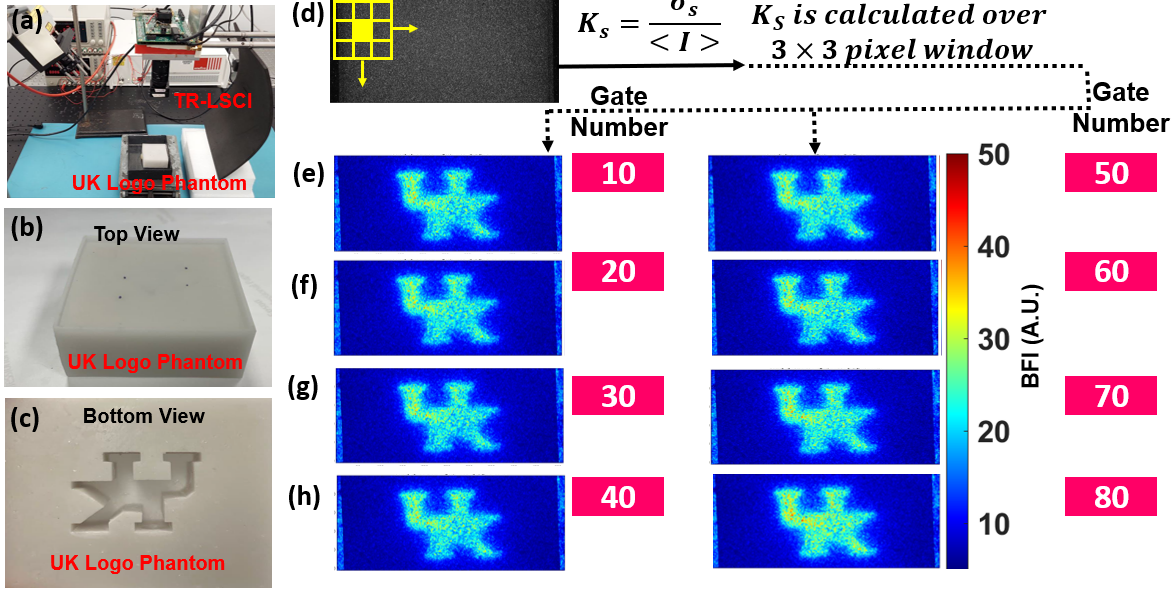}
\caption{Phantom experiments utilizing TR-LSCI in gated mode. (a)-(c) TR-LSCI setup for imaging the UK logo phantom. The 3-D printed solid phantom with the empty UK logo (flow index = 0) were filled in with the Intralipid solution (flow index = 1). (d) Using the LSCI method to calculate Ks. (e)-(f) Resulting 2D maps of Intralipid particle flow contrasts in the phantom with the top layer thicknesses of 1 mm, imaged by the TR-LSCI with the gate numbers ranging from 10 to 80. Images are averaged at each gate to increase the signal-to-noise ratio.} \label{fig:spad_data}
\end{figure*}

\begin{figure*}[bt!]
\centering
\includegraphics[width=1.8\columnwidth]{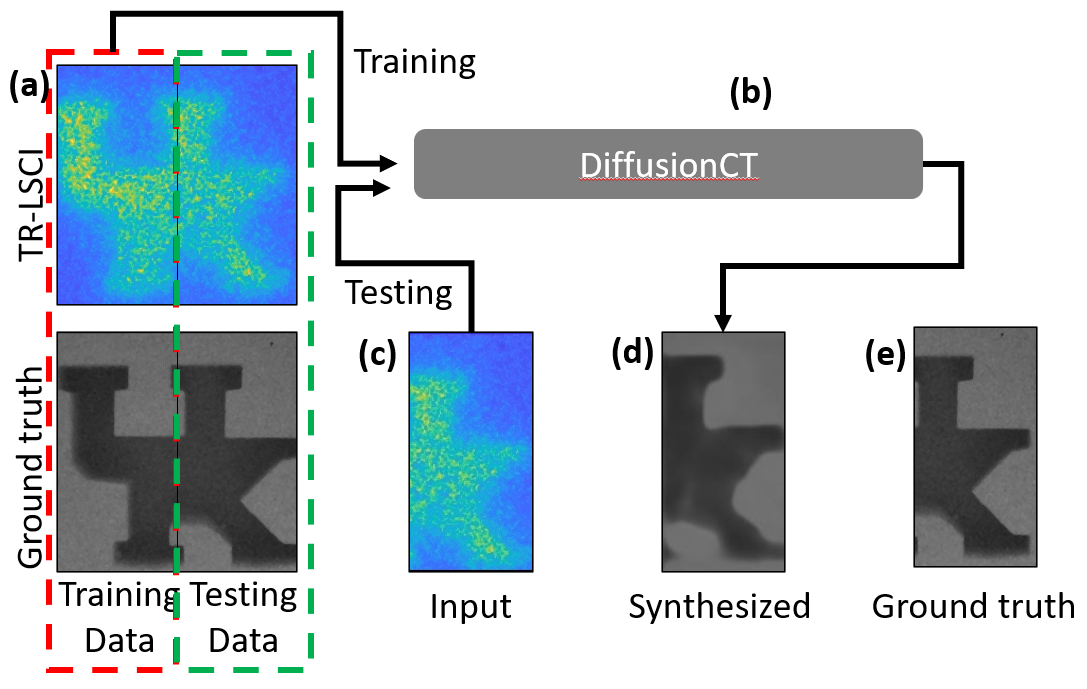}
\caption{DiffusionCT model to reduce photon diffuse noise and results on phantom images. (a) 7,201 image patches extracted from the left side (red box) of the phantom image were used to train the DiffusionCT model (b), and the right side image (green box) were used to test the model. (c)-(e) Testing image, resulting synthesized image, and the corresponding ground truth.} \label{fig:spad_diffusion}
\end{figure*}

Tissue-simulating phantoms with empty channels bearing the University of Kentucky logo (‘UK’) were used to illustrate the fundamental concept of TR-LSCI (Figure~ \ref{fig:spad_data}a-\ref{fig:spad_data}c). The UK phantom consisted of water, Intralipid particles, and India ink (Black India, MA) while the solid phantom was prepared by resin, India ink, and titanium dioxide (TiO2). TR-LSCI illuminates picosecond-pulsed, coherent, widefield near-infrared light (785 nm) onto the phantom and synchronizes a gated single-photon avalanche diode (SPAD) camera to image flow distributions at different depths. See details of TR-LSCI principle and the design of the UK phantom at Fathi et al~\cite{fathi2023time}.


%
%

The SPAD camera’s raw intensity images were taken at the depth of 1mm with different gate numbers. The gated intensity images were then converted to a speckle contrast image based on LSCI analysis: $K_s=\frac{\sigma_s}{<I>}$, where $K_s$ is defined as the ratio of the standard deviation to mean intensity in a pixel window of 3x3 (Figure~\ref{fig:spad_data}d). A flow index can be approximated as the inverse square of the speckle contrast: $BFI \sim 1/K_s^2$. 
Figure~\ref{fig:spad_data}e-\ref{fig:spad_data}f show the results using the TR-LSCI to image the UK logo phantoms. 
These results are expected as deeper penetration and thicker top layer resulted in fewer diffused photons being detected.

DiffusionCT was trained to reduce TR-LSCI image noises. Image with a high noise rate obtained using TR-LSCI was paired with the corresponding phantom shape image (Figure~\ref{fig:spad_data}c) considered the ground truth. Left-half of the phantom image (n=7,201) was used to train the DiffusionCT model and right-half was used to test the model performance (n=7,201). 

Results on the UK logo phantom are shown in Figure~\ref{fig:spad_diffusion}c-\ref{fig:spad_diffusion}e. The resulting image preserves the structural information and contains much less noise than the input. The results were evaluated using the structural similarity index measure (SSIM), concordance correlation coefficient (CCC), and peak SNR (PSNR). The synthesized image (Figure~\ref{fig:spad_diffusion}d), compared to the input image (Figure~\ref{fig:spad_diffusion}c), has improved SSIM from 0.44 to 0.77, PSNR from 12.50 to 23.75, and CCC from -0.01 to 0.86, where all the measurements were computed in reference to the ground truth (Figure~\ref{fig:spad_diffusion}e). 

\section{Conclusion}
Image standardization reduces texture feature variations and improves the reliability of radiomic features of CT imaging. The existing CT image standardization models were mainly developed based on GAN. This article accesses the application DDPM approach for the CT image standardization task. Both image space and latent space have been investigated in relation to DDPM. The experimental results indicate that DDPM-based models are significantly better than GAN-based models. The DDPM has comparable performance in image space and latent space. Owing to its relatively compact size, DiffusionCT is best suited for creating more abstract embeddings in the target domain. 

In this study, we have adopted a ResNET-18-based encoder as it is a widely used CNN architecture. The future research direction includes the comparison with other available architectures, e.g., VGG, and vanilla U-Net. Besides network architecture, the future scope of this study includes experiments with larger and patient datasets.


\bibliographystyle{IEEEtran} 
\bibliography{bibfile} 

\end{document}